\documentstyle[preprint,aps]{revtex}
\def\lsim{\:\raisebox{-0.5ex}{$\stackrel{\textstyle<}{\sim}$}\:}

\def\hroots{{\sqrt{\hat s}\over 2}}
\def\be{\begin{equation}}
\def\ee{\end{equation}}
\def\bea{\begin{eqnarray}}
\def\eea{\end{eqnarray}}

\begin{document}
\draft
\preprint{hep-ph/9402345, TIFR/TH/93-57, BU-TH-93/5}
\title
{\bf High Precision $J/\psi$ and $\Upsilon$-production data 
and the Nuclear Glue} 
\author{R.V. Gavai\footnote{E-mail: gavai@theory.tifr.res.in}}
\address{ Theory Group, Tata Institute of Fundamental 
Research, Bombay 400 005, India\\}

\author{R.M. Godbole\footnote{E-mail: rohini@theory.tifr.res.in}}
\address{ Dept. of Physics, Univ. of Bombay, Vidyanagari, Bombay 400
098, India}
\maketitle

\begin{abstract}
We use the high statistics E-772 data on the nuclear dependence of the
production of quarkonia $(J/\psi$ and $\Upsilon)$ and dimuons at large
transverse momentum $(p_T)$ in $p$-$A$ collisions to get information about
the gluonic EMC effect.  
We find a satisfactory quantitative agreement of the theoretical
predictions with the data although none of the models of the EMC
effect we consider could account for the entire data. Since all the
qualitative features are understood none the less in terms of
perturbative QCD with nuclear dependent parton densities, our results
suggest that these data can now be used for a better determination of
the nuclear parton densities. Our conslusions are shown to be
insensitive to the hadronisation mechanism  for  the quarkonia. 

\end{abstract}

\narrowtext
\newpage

\section{ Introduction} 

The observation of a nontrivial nuclear dependence of the parton
densities, the EMC effect\cite{oldemc}, confirmed later in a further
series of deep inelastic scattering experiments\cite{emcreview}, 
still lacks a clear theoretical understanding even after a decade.  
A large variety of models\cite{emcreview}, with widely differing 
basic underlying physics mechanisms, have been proposed 
to explain the EMC effect.  In all of these, some model parameters 
have to be fitted to reproduce the measured ratio $\rho$ of the
structure function $F_2^{~A}$ (per nucleon) for the nucleus to that of
the nucleon
$F_2^{~p}$. As a result, the nuclear quark densities are very similar in all
these models.  However, their predictions for nuclear gluon densities are
quite different.  Hence a good measurement of the gluonic ratio,
\[
\rho_g = {g^A (x,Q^2) \over A g^P (x,Q^2)}
\]
can play an important role in distinguishing the different models of EMC
effect.  This information can be obtained by studying the nuclear
dependence of different hard scattering processes\cite{roreview} such 
as jet production\cite{souro}, direct photon production\cite{sousri}, 
quarkonium $(J/\psi$ and $\Upsilon)$ 
production\cite{souraj,sriro,souhel,epgaga}, 
$\mu^+\mu^-$ pair production\cite{sulassila} with nuclear targets.  
Recently the associated production of $J/\psi$ and a photon at large
$p_T$  has also 
been suggested\cite{srirajro} as a means to catch a glimpse of the 
gluonic EMC effect and to help unravel the correct model of EMC effect.

Nuclear dependence of quarkonium and $\mu^+\mu^-$ pair production can
be very crucial for the various experimental signals of the formation 
of Quark-Gluon Plasma (QGP) in relativistic heavy ion
collisions.  It has been suggested\cite{Matsat} that due to formation of QGP,
the $J/\psi$ production in $A$-$A$ collisions will be suppressed and the
suppression is expected to exhibit a characteristic $p_T$-dependence with
a maximum suppression at $p_T = 0$\cite{Karsch}.  To analyse
the feasibility of using this suppression as a signal for QGP formation it
is, however, necessary to understand clearly the expectations of
perturbative quantum chromodynamics, (pQCD), for the quarkonia 
production in $A$-$A$ collisions.  The observation of a $p_T$-dependent
$J/\psi$-suppression by the NA38 experiment\cite{NA38} further made such
a study of pQCD imperative.  The applicability of pQCD 
for this process is justified because of the short
time scale for the production of a heavy quark-antiquark $(Q\bar Q)$ pair
which is $O(1/2M_Q)$ for a heavy quark of mass $M_Q$.  In the pQCD
approach the initial state dependence of $J/\psi$ or
$\Upsilon$ production  comes through the nuclear parton densities.  Earlier 
investigations\cite{souraj,sourajsri} have revealed that the nuclear 
dependence of parton densities gives rise to a $J/\psi$-suppression in 
$A$-$A$ and $p$-$A$ collisions which is qualitatively similar to the 
proposed QGP signature.  Since any suppression due to QGP formation or a
hot dense hadron gas amounts merely to a change in the hadronisation
model and/or final state interactions, it is clear that a
critical assessment of the utility of $J/\psi$-suppression as a signal for
QGP formation needs a good understanding of nuclear parton densities.
Since quarkonia are usually detected through their $\mu^+\mu^-$ decay, the
continuum $\mu^+\mu^-$ pair production in $A$-$A$ collisions
has to be investigated as well.  Fortunately, high statistics data on 
$J/\psi $ or $\Upsilon$ and $\mu^+\mu^-$ pair production with nuclear 
targets\cite{E7721,E7722,E7723} has recently become available from the
Fermilab E772 experiment.  In this paper we compute 
large-$p_T$ $J/\psi,\Upsilon$ and dimuon production in $p$-$A$ collisions in
pQCD and compare it with the E772 data with a view to get a handle on
the gluonic EMC effect.  The main advantage of comparing the model
predictions with all these data from the same experiment is of course
in minimizing the effects of systematic uncertainties.
We consider the processes
\[
p + A \rightarrow J/\psi (\Upsilon,\mu^+\mu^-) + X
\]
as due to a hard collision between partons from the proton and the nuclear
target with mass number $A$.  The nuclear dependence is incorporated 
only through the parton densities which is where the different models of
EMC effect come into picture.  The hard scattering cross-sections for
$Q\bar Q$ production $(Q=c$ (b) for $J/\psi (\Upsilon))$ and $\mu^+\mu^-$
pair production are available in literature\cite{hq1,hq2,dimuth}.  
Two different hadronisation models\cite{colsing,sld} exist in the
literature for the hadronisation of the $Q\bar Q$ pair into a quarkonium. 
In our work we use both of them to assess the uncertainty in
our results due to these. Earlier an attempt\cite{souhel}
was made to extract the nuclear glue from the E772 data by 
using the approximation of gluon dominance 
of the partons in the initial state and by assuming that the $Q\bar Q$ 
production takes place at $p_T=0$ for the pair. 
We comment on the reliability of this approach.

The plan of the paper is as follows: in the next section we summarize 
different theoretical models for nuclear parton densities (i.e. the EMC 
effect) which we use. In section III, we first present the details of 
the theoretical formalism we use
to calculate the quarkonia production at large-$p_T$.  We then discuss and
compare the two hadronisation models we employ and point out
the major kinematical differences in the two models.  Following this
the limitations of the gluon dominance approximation,
used in Ref.~\cite{souhel} to extract $\rho_g$ from
$J/\psi$ and $\Upsilon$ production are highlighted.  Finally, we end the
section by presenting a comparison of our pQCD calculations
with the data from the E772 experiment for the $J/\psi$ and 
$\Upsilon$--production at large $p_T$. The theoretical predictions are 
obtained for the different nuclear parton densities by imposing the
experimental cuts of the E772 group on the ranges of the kinematical variables.
We also compare the
predictions of different hadronisation models with each other.  
In the next section, we discuss the formalism for calculating
the background, continuum $\mu^+\mu^-$ pair production at large-$p_T$ and
present our results for the ratio of partially integrated cross-sections of
dimuon production for different nuclear targets and parton densities and
compare them with the E772 data.  We then present our conclusions in the
last section.

\section{ Models for nuclear parton densities}

We define the nuclear parton densities by removing a factor $A$, where $A$
is the atomic number of the nucleus, so that the cross section defined
below  is per nucleon of the target,
making it directly comparable to the $pp$ case.  The presence of the
EMC-effect tells us that the nuclear parton densities are different 
from those of a nucleon even after this rescaling, i.e., the ratio
$\rho$, defined in the introduction, is a nontrivial function
of $x$, where $x$ is fraction of the momentum of a $nucleon$ which
the parton carries: $0 \le x \le A$ for the nucleus.  There are
many models in the literature which predict $\rho$ in terms of
some parameters which can be fixed by comparing with the data on
the EMC-effect.  The corresponding $\rho_g$ is then usually
predicted in these models.  We limit the
scope of our present investigations by
concentrating on three different models, which are chosen as typical
examples of sets of models based on similar physical concepts:
i) the gas model\cite{gas}, ii) the rescaling model\cite{resc} and
iii) the six-quark cluster model\cite{bag}.  In all the three cases,
we neglect QCD-evolution, since such corrections
are small in the kinematic range of interest to us.  Fig. \ref{fig1}
displays the predicted $\rho_g$ for all these three models.  The details
of these predictions are given below for each of the model.

\subsection{The Gas Model} 

The parton densities of a nucleus with atomic number $A$ are defined in the
Gas model\cite{gas} as a sum of two components.

\be 
f_{i/A}(x) = (1 - \omega) {\tilde f}_{i/N}(x) + {1\over A}
\sum^A_{r=1} \omega^r (1-\omega)^{A-r} f_{i,r}^{\rm gas} (x;\mu,T)
\label{gasmodel}
\ee

The first term, occuring with a weight $(1 - \omega)$, is that for a
free nucleon parton density after corrections for its Fermi motion inside
the nucleus.  The second component is written in terms of thermal
distributions of momenta at a temperature $T$,
leading to the following functions $f_r^{\rm gas} (x;\mu,T)$:
\be
f_{q,r}^{\rm gas} (x;\mu,T) = {2r_0^3 T^3 \over \pi} \Bigg[\phi ~\ell n(1 +
ze^{-\phi}) + {1\over2} \ell n^2(1 + ze^{-\phi}) + Li_2\bigg({z \over
z+e^\phi}\bigg)\Bigg]
\label{gasq}
\ee
and
\be
f_{g,r}^{\rm gas} (x;\mu,T) = {2r_0^3 T^3 \over \pi} \bigg[Li_2(e^{-\phi}) -
\phi ~\ell n(1 - e^{-\phi})\bigg]
\label{gasg}
\ee
Here $r_0 = 1.2$ fm, $\phi = Mx/2T$, $z = {\rm exp}(-\mu/T)$, where $M$
is the nucleon mass and $Li_2(x)$ is the Euler dilogarithm function.
Using the constraint on the total baryon number of the nucleus to
eliminate the chemical potential $\mu$, one has two model parameters,
$T$ and $\omega$, for each nucleus.

Using the CDHS parametrisations\cite{cdhs},
\bea
\label{cdhs}
 F_2^{\nu p}(x)\;&=&\;1.1(1+3.7x)(1-x)^{3.9}\nonumber\\[2mm]
 x\sigma_p(x)\;&=&\;0.17(1-x)^{8.54}\\[2mm]
 xf_{g/p}(x)\;&=&\;2.62(1+3.5x)(1-x)^{5.9}\nonumber
\eea
for $F_2^p(x)$ (here $\sigma_p(x)$ is the total sea density) and 
the data on $\rho(x)= F_2^A(x)/A F_2^p(x)$
these parameters have been fixed\cite{gas} for many nuclei, including the 
ones used for the E772 experiment.  The corresponding $\rho_g (x)$ 
is then predicted uniquely. We summarise them in Table \ref{tabgas}.
It may be mentioned here that using a different
set of structure functions for proton instead of eq.\,(\ref{cdhs})
necessitates a time-consuming re-analysis of the EMC-data to obtain 
$T$ and $\omega$.  For this reason we have not used more recent 
parametrisations for the proton structure function and we are also
constrained to use {\em different} proton structure functions in
different models.  

\subsection{ The Rescaling Model}

Scaling models\cite{resc,GGB,G} seek to explain the EMC effect as
arising from the change in the QCD-scale in going from a nucleon to a
nucleus.  The nuclear parton densities at a scale $Q^2$ are obtained
from the parton densities in a proton at the same $Q^2$ by evolving
them to a scale $\xi_A Q^2$, i.e., the nuclear parton density per
nucleon $f_{i/A} (r, Q^2)$ is given by
\be
f_{i/A}(x, Q^2) = f_{i/p} (x, \xi_A Q^2)
\ee
In this paper, we use the rescaled nuclear densities as obtained in
Refs. \cite{GGB,G} where $\xi_A = A^{2/3}$ and the starting nucleon
parton densitites were taken\cite{G} to be a parameterisation 
of the EMC Deuterium data at $Q^2 = 20~{\rm GeV}^2$.  For further
details, we refer the reader to Ref. \cite{G}.

\subsection{ The six-quark cluster model}

The six-quark cluster model\cite{bag} is also a 
representative of the two-component models for EMC effect.
In this model, it is assumed that when two nucleons get closer
to each other than a certain critical radius they merge together to 
form a six-quark cluster.  By assuming the probability to form higher
clusters to be negligible, the remaining model inputs are the probability of
forming such a cluster and the form of the parton distributions in the
3-quark and 6-quark clusters.  The latter are chosen using the
quark-counting rules and the constraints of i) 
normalization of valence densities (an N-quark cluster has N valence
quarks), and ii) conservation of momentum.  It is further assumed that the
average momentum carried by the sea partons is the same
for the three- and six-quark clusters and 
that it is $\sim 0.2$ of that of the gluons.  The forms of
nuclear densities per nucleon in this model are,
\be
f_{i/A} (x) = (1 - \epsilon) f_{i,3} (x)
+~ {\epsilon \over 2}~f_{i,6} \left({x \over 2}\right)~~,~~ 
\ee
where $\epsilon$ is the probability to find a six quark cluster which
increases with $A$ \cite{pirner} and the subscripts denote the cluster
size.  The values of $\epsilon$ which we used 
are given in Table \ref{tabgas}.  Specific choices\cite{bag} for 
the valence density $V(x) = f_{u_V}(x) + f_{d_V}(x)$, sea density $S(x)$ and
the gluon $G(x)$ for an $N$ quark cluster ($N = 3, 6)$, which we
used in our analysis presented here, are given by
\bea \label{bagden}
x V_N (x)&=&N x^{0.5} (1-x)^{2N-3} \biggm/ B\left(\frac{1}{2},~2N -2
\right) \nonumber\\[2mm]
x S_N (x)&=&{N-1 \over 2(4N - 3)}~ (a_N + 1) ~(1 -x)^{a_N} \\[2mm]
x f_{g,N}(x) \equiv x G_N(x)&=&{5(N - 1) \over 2(4N-3)} ~ (c_N +1)
~(1-x)^{c_N}~~,~~\nonumber
\eea
with $a_3 = 9,~ a_6 = 11, ~~ c_3 = 7$, and $c_6 = 10$\cite{sulassila}.
Here $B$ is the usual Euler function and
$S_N(x)$ represents the sum of sea quark densities over all flavours.
With a further assumption of 
$\bar f_{\bar s,N} (x) = {1 \over 2}~f_{\bar u,N} (x) = {1 \over 2}
~f_{\bar d,N} (x)$,
the $\bar u$ distribution for the $N$-quark cluster is given by
$f_{\bar u,N} (x) = {1 \over 5}~ S_N (x)$.

\section{ Theoretical formalism }

The problem of heavy quark-antiquark pair $(Q\bar Q)$
production has been studied extensively both experimentally and
theoretically.  In the pQCD approach 
the quarkonium production cross-section can be
calculated by convoluting the hard scattering subprocess cross-section 
(at a give order in $\alpha_s$) with the appropriate
initial parton densities and suitably chosen hadronization functions.  
Two popular hadronisation models which describe conversion of the $Q\bar
Q$ pair into quarkonia  are i) the semilocal-duality
(SL) model and ii) the colour singlet (CS) model.  Since we employ both of 
them in our work, we will discuss them briefly below.  As we shall see,
the differences in their details result in different kinematics for the
same process at formally the same order of perturbation theory.  It is
therefore interesting to compare their predictions with the E772 data to
check the robustness of the pQCD approach; any differences are likely to
provide a clue on the hadronisation of the $(Q\bar Q)$-pair.

\subsection{ Semilocal Duality Model}

In this model one first computes the basic $Q\bar Q$ production
cross-section. The quarkonium cross-section is then obtained by simply
restricting the invariant mass of the $Q\bar Q$ pair between $2M_Q$ and
$2M_{h_Q}$ (where $M_Q$ is the heavy quark mass and $M_{h_Q}$ is the mass
of lowest lying $q$-flavoured meson) and by multiplying the cross-section by
a constant.  The underlying assumption is that irrespective of its
invariant mass the $(Q\bar Q)$-pair is equally likely to turn in to a 
given quarkonium provided it is below the threshold of the heavy 
flavoured mesons.  The constant {\em may} depend on the colliding energy.
However, we will need to consider only ratios of cross-sections for the
nucleon and nuclear target at a fixed energy, and the constant drops
out of such ratios.
The $Q\bar Q$ cross-section itself is given by,
\bea
\sigma(pA \rightarrow Q\bar Q X)  =  \sum_{p_1,p_2} 
\int^1_{4M^2_Q \over s} dx_1 \int^1_{4M^2_Q \over sx_1} dx_2 
& & \left[f_{p_1/p} (x_1) f_{p_2/A} (x_2) + f_{p_2/p} (x_1) f_{p_1/A}
(x_2)\right]\nonumber\\[2mm]
& &\hat \sigma_{p_1p_2} (\hat s,M^2_Q)
\label{one}
\eea
Here $f_{p/h} (x)$ is the parton density for parton $p$ in hadron $h$, $x$
is the momentum fraction of $h$ carried by $p$ and $\hat \sigma$ is the
appropriate subprocess cross-section integrated over all the subprocess
variables.  The square of the total  centre of mass (cm) energy of the
partonic system is given by
\[
{\hat s} = {sx_1 x_2}
\]
where $s$ is the square of the cm energy of the $p$-$A$ system.

For $Q\bar Q$ production both the $O(\alpha^2_s)$\cite{hq1} and
$O(\alpha^3_s)$\cite{hq2} expressions for $\hat \sigma$ are available.  
The subprocesses which contribute at $O(\alpha^2_s)$ are
\be
gg \rightarrow Q\bar Q, ~~~q\bar q \rightarrow Q\bar Q
\label{two}
\ee
and the ones contributing at $O(\alpha^3_s)$ are
\be
gg \rightarrow Q\bar Qg, ~~~q\bar q \rightarrow Q\bar Qg, ~~~qg
\rightarrow Q\bar Qq, ~~~\bar qg \rightarrow Q\bar Q\bar q
\label{three}
\ee
The quarkonia produced from $O(\alpha^2_s)$ subprocess can have only small
$p_T$, being of the order of the intrinsic transverse momentum of the 
incoming partons.  On the other hand, the $O(\alpha^3_s)$
processes will yield large $(\sim O$(GeV)) $p_T$ for the quarkonium
with the light parton in the final state constituting a jet .  Thus
the $J/\psi(\Upsilon)$ produced at large-$p_T$ in the $SL$ model come from
$2 \rightarrow 3$ subprocesses.

The matrix elements squared, appropriately averaged, for these processes have
been computed. There are clearly two classes of diagrams---those
involving one gluon, and those with three. We shall denote the squared
matrix elements by $A$ and $B$ respectively. They depend on the
momenta of all the five particles. Instead of reproducing the lengthy
expressions for them, we refer the reader to the original work
\cite{ellis}. For computing $p_T$-distributions of the quarkonia
we will need cross-sections more differential than the expression in
eq.\,(\ref{one}). 

The phase-space for these processes is 6-dimensional.
We consider the kinematics in the EHLQ \cite{ehlq} conventions, modified
to account for massive $Q$ and $\bar Q$.
Defining the parton momenta in their center of mass, one sees that the three
final state particles lie in a plane. We choose this to be the $xy$-plane, 
and orient our axes by choosing the jet to be in the positive
$x$-direction. The 4-momenta of the $Q$, $\bar Q$ and the jet are
given, respectively, by
\bea
p_3\;&=&\;\hroots~x_3(1, \beta_3\cos\theta_{35}, 
                  \beta_3\sin\theta_{35},0),\nonumber\\[2mm]
p_4\;&=&\;\hroots~x_4(1, \beta_4\cos\theta_{45}, 
                  \beta_4\sin\theta_{45},0),\\[2mm]
p_5\;&=&\;\hroots~x_5(1,1,0,0), \nonumber
\label{fspmomenta}
\eea
where $ \beta_i=\sqrt{1-4M_Q^2/ x_i^2s}$ and the angles are given by
\be
 \cos\theta_{35}\; =\;{( \beta_4 x_4)^2-( \beta_3 x_3)^2
      - x_5^2\over2 \beta_3 x_3 x_5}
     \quad{\rm and}\quad
 \cos\theta_{45}\; =\;{( \beta_3 x_3)^2-( \beta_4 x_4)^2
                         - x_5^2\over2 \beta_4 x_4 x_5}.
\ee
Furthermore, energy-momentum conservation implies $x_3 + x_4 + x_5 = 2$,
($0\leq x_i\leq1$), leaving only two independent variables in the
equations above. In addition to $x_1$ and $x_2$ of eq.\,(\ref{one}),
two more variables are needed to specify the momenta
of the incoming partons in this frame. Choosing these to be the
Euler angles $(\theta,\phi)$, we get
\bea
\label{ispmomenta}
p_1\;&=&\;\hroots~(1,-\sin\theta\cos\phi,
          -\sin\theta\sin\phi,-\cos\theta),\nonumber\\[2mm]
p_2\;&=&\;\hroots~(1,\sin\theta\cos\phi,\sin\theta\sin\phi,\cos\theta).
\eea
We can eliminate $\theta$ in favour of the transverse momentum of the
pair (which is equal to that of the jet) through the relation
\be 
p_T\;=\;\hroots~x_5\sqrt{\cos^2\theta+\sin^2\theta\sin^2\phi}.
\label{ptdef}
\ee
Similiarly we can write the Feynman scaling variable $x_F$ for $J/\psi$ as
\be
x_F\;=\;{1\over2}\left[(x_1+x_2)x_5\sin\theta\cos\phi
          +(x_2-x_1)(x_3+x_4)\right].
\label{xfdef}
\ee
The E772 data has cuts on this $x_F$ which we include in our computations.
 
Using the above kinematic relations the fully
differential cross section for $J/\psi$-production computed to order
$\alpha_s^3$ can be written down as below:
\bea
\label{thirdcs}
 & &{d\sigma_{pA}\over dx_1dx_2dx_3dx_4d\phi dp_T}\;=\;
    {\alpha_s^3 p_T\over16\pi s x_1 x_2 x_5^2
             \cos\theta\cos^2\phi}\nonumber\\[2mm]
 & &\qquad  \Biggl[{1\over9}\sum_q\left\{f_{q/p}(x_1) f_{\bar q/A}(x_2)+
                    f_{\bar q/p}(x_1)f_{q/A}(x_2)\right\}
                        A(p_3,p_4,-p_1,-p_2,p_5) \nonumber\\[2mm]
& &\qquad        -{1\over24}\sum_q\left\{f_{g/p}(x_1)f_{q/A}(x_2)+
                   f_{q/p}(x_1)f_{g/A}(x_2)\right\}
                        A(p_3,p_4,p_5,-p_1,-p_2)\\[2mm]
 & &\qquad        -{1\over24}\sum_q\left\{f_{g/p}(x_1)f_{\bar q/A}(x_2)+
                   f_{\bar q/p}(x_1)f_{g/A}(x_2)\right\}
                        A(p_3,p_4,-p_1,p_5,-p_2)\nonumber\\[2mm]
 & &\qquad        +{1\over64}f_{g/p}(x_1)f_{g/A}(x_2)
                        B(p_4,p_3,p_5,-p_1,-p_2)\Bigg].\nonumber
\eea       
This is the cross section formula which together with appropriate structure
functions, chosen from section II, we integrated over the kinematic region 
corresponding to the E772 experimental cuts\cite{E7721,E7722} to compute 
the $p_T$ distributions for the quarkonia $J/\psi$ and $\Upsilon$.

\subsection{ Colour Singlet Model}

The colour singlet model was first developed for photoproduction of
quarkonia\cite{bj} and later generalised to the hadronic 
production\cite{colsing}.  In this model one projects out the 
state with appropriate spin, parity and
colour assignments from the full $Q\bar Q$ production amplitude to match
the quantum numbers of the resonance under consideration.  The projection
is done at the level of hard scattering amplitude itself and this yields a
multiplicative factor related to the quarkonium wave function at the
origin in co-ordinate space.  The effect of the hadronisation of the
$Q\bar Q$ pair into the quarkonium is thus contained in this factor.  For
an $S$-wave resonance this multiplicative factor is the wave function,
$R(0)^2$, at the origin whereas for a $P$-wave resonance it is the
derivative of the wave function at the origin, $R'_1(0)^2$.  $R(0)$
is related to the measured, leptonic $^3S_1$ width by
\be
\Gamma_{\ell\bar\ell} (^3S_1) = {4\alpha^2 e^2_Q R(0)^2 \over M^2}
\label{five}
\ee
where $\alpha$ is the electromagnetic coupling, $e_Q$ is the quark
charge in units of proton charge and $M$ is the mass of the quarkonium.
$R'_1 (0)$ can be related to the
total hadronic width of the resonance by assuming it to be approximately
the same as its gluonic width given by,
\be
\Gamma_{gg} (^3P_0) = {96 \alpha^2_s R^{\prime 2}_1 (0) \over M^2} 
\label{six}
\ee
with $\alpha_s$ being the running strong coupling.  

The model is known to give a good description of the kinematical 
distributions in leptoproduction\cite{EMCNMC} and
hadroproduction\cite{colsing,Albajar} of $J/\psi$.  However, there is a 
considerable uncertainty in the overall normalisation.  
Even after using the QCD corrected version of eq.\,(\ref{five}) the 
data\cite{EMCNMC} required a $K$-factor of 2.4.  This large $K$-factor 
is perhaps due to the nonrelativistic treatment of the quarkonium 
$J/\psi$ in arriving at the hadronisation factor $R(0)^2$.
Once again, for ratios of cross-sections at the same colliding energy,
which we will consider here, the precise value of $K$-factor plays no 
role; we assume it to have no nuclear dependence.

The quarkonium production cross-section for the $^{2S+1}L_J$ quarkonium
state in this model is given by
\bea
\label{seven}
\sigma^{CS} (p + A  \rightarrow ~^{2S+1}L_J + X) = 
&\sum_{p_1,p_2,p_3} \int^{p_T^{\rm
max}}_{p_T^{\rm min}} dp_T \int^{y_1^{\rm max}}_{y_1^{\rm min}} dy_1
\int^{y_2^{\rm max}}_{y_2^{\rm min}} dy_2 ~2p_T x_1 x_2 \nonumber \\[2mm]
&\times \left[f_{p_1/p} (x_1) f_{p_2/A} (x_2) + f_{p_2/p} (x_1) f_{p_1/A}
(x_2)\right] \nonumber \\[2mm]
&\times d\hat\sigma/d\hat t (p_1 + p_2 \rightarrow ~^{2S+1}L_J + p_3) 
\eea

Here $x_1,x_2$ again denote the momentum fractions of the proton and
nuclear target respectively carried by the partons $p_1,p_2,f_{p_i/h}$ are
the parton density distribution functions and $d\hat \sigma/d\hat t$
denotes the differential subprocess cross-section.  Here $y_1,y_2$ denote
the rapidities of the quarkonium and the jet respectively and $p_T$ is the
transverse momentum of the quarkonium.  The variables $x_1,x_2$ are given
in terms of $y_1,y_2$ by
\bea
x_1 &=& {1 \over 2} \left[\bar x_T e^{y_1} + x_T e^{y_2}\right] \nonumber
\\[2mm] x_2 &=& {1 \over 2} \left[\bar x_T e^{-y_1} + x_T e^{-y_2}\right]
\label{eight}
\eea
In the above equation $x_T = {2p_T \over \sqrt{s}}$, $\bar x_T =
\sqrt{x^2_T + 4\tau}$ with $\tau = M^2/s$.  The total allowed
range of integration over $y_1,y_2$, for a given value of $p_T$ is given by
\bea
|y_1| &\leq& \cosh^{-1} \left({1 + \tau \over \bar x_T}\right) \nonumber
\\[2mm] - \ell n\left({2 - \bar x_T \exp(-y_1) \over x_T}\right) &\leq&
y_2 \leq \ell n\left({2 - \bar x_T \exp(y_1) \over x_T}\right)
\label{nine}
\eea
$\hat s,\hat t,\hat u$ are the Mandelstam variables of the subprocess
given by 
\[
\hat s = x_1 x_2 s;~~
\hat t = M^2 - x_1 \sqrt{s(p_T^2 + M^2)} e^{-y_1};~~ 
\hat u = - x_1 p_T \sqrt{s}  e^{-y_2}. 
\]

At the lowest order in the strong coupling, $O(\alpha^2_s)$,
there is no parton $p_3$ in the final state; one has only the gluon 
fusion process
\be
gg \rightarrow ~^1S_0, ~^3P_{0,2}
\label{ten}
\ee
The quarkonium so produced has $p_T \simeq 0$.  At $O(\alpha^3_s)$, the $gq
(g\bar q)$ and $q\bar q$ scatterings give rise to $^1S_0$, $^3P_J$ 
resonances ($\chi $ states)  via
\bea
gq (\bar q) &\rightarrow& ^1S_0, ~^3P_J + q(\bar q); \nonumber \\[2mm]
q\bar q &\rightarrow& ^1S_0, ~^3P_J + g
\label{eleven}
\eea
while the $^3S_1$ resonance can be produced directly as well
via the $gg$ subprocess,
\be
gg \rightarrow ~^1S_0, ~^3S_1, ~^3P_J + g
\label{twelve}
\ee
The $^3P_J$ states can decay into $^3S_1$ states via the radiative decay
\be
^3P_J \rightarrow ~^3S_1 + \gamma~~,
\label{thirteen}
\ee
thus giving to rise to indirect contribution to the
production of $~^3S_1$ in addition to that of eq.\,(\ref{twelve}).
The expressions for the differential cross-section for the various
subprocesses are available\cite{colsing,wu} and are not reproduced here.
 
Note that according to this model $^3P_1$ production is not possible at
$O(\alpha^2_s)$ and hence the total $^3P_1$ rates are expected to be small
compared to that of $^3P_0,^3P_2$.  However this seems to be belied by the
data\cite{Antoniazzi} on hadroproduction of $^3P_J$ charmonium states.  
Since soft gluons, ignored as a part of the model ansatz, are likely
to be more relevant in this case, this possibly
points to problems with the $CS$ model for the $p_T \simeq 0$
quarkonium production.  However, we use it here only for large-$p_T$
($\sim O$ (GeV)) quarkonium production where the soft gluons most
likely do not play such a significant role.

We compute the $^3S_1$ $J/\psi(\Upsilon)$ production cross-section including
both the direct contribution of eq.\,(\ref{twelve}) and the contribution 
from the decay of $^3P_J$ states where $^3P_J$ state cross-sections are 
computed using eqs.\,(\ref{eleven}) and (\ref{twelve}).  Since the 
$\Upsilon$ produced in process in eq.\,(\ref{thirteen}) carries very 
little energy, we assume the kinematical variables
for the $^3P_J$ and $^3S_1$ resonance to have essentially the same values. 
The ${d\sigma / dp_T}$ is obtained using eq.\,(\ref{seven}).  
The E772\cite{E7721,E7722} cuts on the Feynman $x_F$ of the resonance 
were implemented by restricting the $y_1$-integration between $y_1^{\rm min}$
and $y_1^{\rm max}$, which are related to the allowed range of $x_F$,
$x^-_F < x_F < x^+_F$, by
\bea
y^{\rm min}_1 = \ell n \left({x^-_F \over \bar x_T} + \sqrt{\left({x^-_F
\over \bar x_T}\right)^2 + 1} \right) ~~~~~\nonumber \\[2mm]
y_1^{\rm max} = \ell n\left({x^+_F \over \bar x_T} + \sqrt{\left({x^+_F
\over \bar x_T}\right)^2 + 1}\right) ~~.~~
\label{fourteen}
\eea
Tables \ref{tabres1} and \ref{tabres2} give the details of  
the set of the resonance parameters which we used for our
computations. The masses of all the resonances are taken from the
latest compilation of particle properties \cite{pdg}.
Note here that the value of the $R(0)^2$ for the S--wave resonances 
$\psi$ and $\psi^\prime$ are determined from the updated measurement
of the letonic decay width \cite{pdg} while the $R(0)^2$ for the 
$\Upsilon$ system and $R_1^\prime(0)^2$ values for both the $\psi$ and
the $\Upsilon$ system have been taken from  ref.\cite{colsing}.

\subsection{ Comparison of the Hadronisation Models}

From the above discussion, it should be clear that the two
hadronisation models differ quite significantly in the time scale for
quarkonium formation.  In the $CS$ model it is the same as the
perturbative time scale $(\sim 1/2M_Q \simeq 0.07$ fm for $Q=c$) whereas
in the SLD case it is a typical hadronic scale $(\sim 1~fm)$.  As a
result, the only effect that a medium like the Quark--Gluon--Plasma
(QGP) can have in the first case will be on
the propagation of the quarkonium,  whereas in the SLD picture the formation
process itself can be affected by the QGP environment.

The kinematics of the two models is also quite different.  In the CS case
basic hard subprocess which gets convoluted with the parton densities (as
in eq.\,(\ref{seven})) is a $2 \rightarrow 1$ subprocess for 
$O(\alpha^2_s)$ case
and $2 \rightarrow 2$ subprocess for $O(\alpha^2_s)$ case of large-$p_T$
quarkonium production.  In the SLD case, these are respectively a
$2 \rightarrow 2$ subprocess and a $2 \rightarrow 3$
subprocess.  Hence, in principle the
momentum division between the large-$p_T$ quarkonium and the light parton
jet can be quite different in the two cases and it is worthwhile to find
out whether perhaps the $p_T$ distribution $d\sigma/dp_T$ for quarkonium
could help discriminate between these two models of hadronisation.

\subsection{ Gluon dominance} 

In Ref.\cite{souhel} it was argued that the E772-data are described by
$O(\alpha^2_s)$ partonic cross-section and further that this
$O(\alpha^2_s)$ production cross-section of eq.\,(\ref{one}), is dominated by
the $gg \rightarrow Q\bar Q$ contribution of eq.\,(\ref{two}).  Then the
ratio of experimentally measured cross-sections for different targets, at
a given value of $x_F$ and $\tau$  directly yields 
the ratio of gluon densities for the two targets at an $x_2$ given by 
\be
x_{1,2} = {1 \over 2} \left(x_F \pm \sqrt{x^2_F + \tau^2}\right)~~.~~
\label{sixteen}
\ee
It is worth noting that only for the case of $2 \rightarrow 1$ kinematics the
$x_F$ of the quarkonium is related to $x_1$ and $x_2$ through the
oft-used simple relation of eq.\,(\ref{sixteen}).  As can be seen from 
eq.\,(\ref{xfdef}), this is no longer true for quarkonium production at
large-$p_T$ via the $2 \rightarrow 3$ subprocess of SLD or also the
$2 \rightarrow 2$ subprocess in CS-model.  Considering the rather large
$p_T$-range of the E772-data $(\leq 2.25$ GeV for the $J/\psi$ and
$\leq 4 $ GeV for the $\Upsilon$), it seems to be an
oversimplification to employ either the eq.\,(\ref{sixteen}) or indeed
the $O(\alpha_s^2)$ subprocess itself (unless one tolerates rather large 
values of intrinsic $k_T$ for the partons) .  As mentioned above, 
since the CS-model is perhaps unreliable for $p_T \sim 0$, even
if eq.\,(\ref{sixteen}) were to be valid one still further needs the 
assumption of gluon dominance if the SLD-model of hadronisation is employed.  

In support of their argument of gluon dominance 
Ref. \cite{souhel} considered the ratio of parton luminosities,
\be
R_{qg} = {F_{qq} \over F_{gg}} = {\sum_{q=u,d,s} \big( f_{q/h}(x_1) 
f_{\bar q/h}(x_2)  + f_{\bar q/h}(x_1) f_{q/h}(x_2) \big) \over 
f_{g/h}(x_1) f_{g/h}(x_2)}
\label{seventeen}
\ee
for a particular parametrisation of the parton densities in the proton 
viz. DFLM\cite{DFLM} and showed it to be rather small in the $x_F$-range
of interest (using eq.\,(\ref{sixteen})).  It turns out, however, that
this demonstration of gluon-dominance strongly
depends on the choice of parton densities used.  We
plot in Fig.\ref{fig2} the ratio in eq.\,(\ref{seventeen}) as a 
function of $x_F$ for $\sqrt{\tau} = 0.0775 $ corresponding to 
the nucleon-nucleon c.m. energy of the E772 experiment and 
$M_{Q\bar Q} = M_{J/\psi}$.
$x_1$ and $x_2$ for a given value of $x_F$ are obtained from
eq.\,(\ref{sixteen}).  The different curves of Fig.\ref{fig2} 
correspond to various popular choices of the parton densities: the older 
parametrisations DFLM\cite{DFLM}, DO1, DO2 \cite{DO} and GHR \cite{GHR}, 
and the newer MT1, MT2, MT3\cite{MT} and ON\cite{ON}.  For the 
DFLM parametrisation which the authors of Ref.\cite{souhel} use
the approximation of gluon dominance is indeed good for low values 
of $x_F$ $(x_F \lsim 0.3)$ .  However, E772 data goes upto $x_F \simeq
0.65$.  Moreover, we also see from Fig.\ref{fig2} that gluon dominance 
crucially depends on the choice of parton density parametrisation in a proton.
For the MT parametrisation, e.g., the $q\bar q$ contribution is $\sim 60$\% of
the $gg$ contribution at the largest $x_F$ value considered $(x_F = 0.5)$.
As Fig.\ref{fig3} shows the situation becomes much worse for $\Upsilon$ with
$\sqrt{\tau} = M_\Upsilon/\sqrt{s} = 0.2365$.  Thus it is clear that the
extraction of $\rho_g(x)$ by using gluon dominance and then simply
taking the ratio of experimentally measured $J/\psi$-cross sections
for the nuclear and the nucleon target has an intrinsic uncertainty
of $\sim10$-15 \%.  In view of the fact that the expected deviation
of $\rho_g$ from unity, viz. the EMC effect, is also of the same order
of magnitude, we feel that this is not a very effective or precise
determination of $\rho_g$.

The models for EMC effect always give a parametrisation for nuclear parton
densities for a specific parametrisation of parton densities in proton.  In
all the models we use, the choice of reference parton densities in the
proton is different from all the above mentioned parametrisations.  We
have checked that our observation about lack of the gluon dominance holds
for these parametrisations as well.

In view of the above discussion, attempts to extract the nuclear gluon
density from the high statistics E772 data appear to have both
conceptual problems and sizeable theoretical uncertainties.  An alternative, 
albeit a less ambitious, approach may be to check the
consistency between the predicted ratios of differential cross-sections
$d\sigma/dp_T$ for various models of nuclear parton densities and the
high statistics E772 data.  One may thus hope to pin them down or even expose
their deficiencies using the high quality and more differential data.
Since the small-$p_T$ region is plagued by the issues of intrinsic
transverse momentum of the inital partons or resummations of higher
order diagrams, we choose to investigate only the large-$p_T$ E772 data 
and compute only the $O(\alpha^3_s)$ contributions to the cross-sections.
This compels us to ignore the data on $p_T$-integrated $x_F$ distributions 
or their ratios since they should receive substantial contribution from
the small-$p_T$ region.  In order to test the $x_F$-dependence of our
pQCD calculations, it would be desirable to have the data integrated
in different, at least two, ranges of $p_T$.

\subsection{Results}

The E772 experiment has provided data for the ratio
\be
R^{J/\psi}(p_T) = {d\sigma(pA \rightarrow J/\psi X) \over dp_T} \biggm/
                   A {d\sigma(pp \rightarrow J/\psi X)\over dp_T}
\label{rjpsiexp}
\ee
with an $x_F$-cut of $0.15 \le x_F \le 0.65$ on the $J/\psi$'s, while
for the $\Upsilon$-production cross sections, they chose to present only
$\alpha(p_T)$, where
\be
 {d\sigma(pA \rightarrow \Upsilon X) \over dp_T} = A^{\alpha(p_T)}
 {d\sigma(pp \rightarrow \Upsilon X) \over dp_T} 
\label{rupsexp}
\ee
with a corresponding $x_F$-cut for $\Upsilon$ of $-0.2 \le x_F \le 0.6$.
The nuclei used were carbon, calcium, iron and tungsten.  Incorporating
these $x_F$-cuts, using eqs.\,(\ref{xfdef}) and (\ref{fourteen}),
we computed each of the individual $p_T$-distribution in 
eqs.\,(\ref{rjpsiexp}-\ref{rupsexp}) for both the SLD-model and the CS-model
and the three models of nuclear structure functions discussed in sect. II.
Fig. \ref{jpsisld} exhibits our results for the SLD-model for all the four
nuclei along with the corresponding data from the E772 collaboration.
The errors for theoretical predictions are purely statistical, arising from
the Monte Carlo integration of the differential cross sections.  
One sees that for the lighter nuclei both the two-component models, namely,
the gas model and the six-quark cluster model, describe the data well, 
especially if one takes into account a possible systematic error of a few
per cent due to variations in input parameters such as $M_c$, $\Lambda_{QCD}$
etc.   For the tungsten nucleus, however, {\em none} of the models seems to
to be in agreement with data. One could compare the CS--model results
with the E772--data in a similar manner as well. Instead we choose to
compare the results of the two hadronisation models directly since the
theoretical results have a better precision.   Fig. \ref{jpsirat}  
shows the ratio of $R^{J/\psi} (p_T)$ for the SLD-model and the 
CS-model for each model of the nuclear parton densities, 
nucleus and $p_T$.  The ratio of the ratios seems to lie in almost all the
cases within $\sim5\%$ of unity.  Considering the different kinematics
of the two models and also the different physics of the quarkonium formation
the agreement is really remarkable and it shows the robustness of the
pQCD predictions.   Note that even in the case of Tungsten both the models
agree with each other rather well and thus have essentially the same
discrepancy with the E772 data.
Of course, the discrepancy in the case of tungsten does
expose the inadequacy of all the three models of the EMC effect and
the corresponding parametrisation of the nuclear parton densities
considered here but the generalagreement in other cases, 
on the other hand suggests  that the structure function effects
indeed do describe the bulk of the $p_T$-data.  One may note here that
the lowest $p_T$ value at which we performed these computations is somewhat
low, being 0.79 GeV.  Presumably, the anticipated large
QCD-corrections at such low $p_T$ 
affect all the cross sections similarly and thus cancel out in the ratio,
resulting in a good description of the data.
A remark on the gluon dominance may be in order as well. Extracting the
contribution for individual parton subprocesses, we typically found the
dominant contribution to be from $gg$ and $qg$ processes which were in the
range of 75-80\% and 25-20\% respectively.  Thus, the quark contributions
are sizeable and an independent determination of the gluon density from
these data are not possible without making an ansatz for the quark 
distributions.

Fig. \ref{upsalph} shows a comparison of our calculations for the SLD-model
for $\Upsilon$-production with the E772-data.  At each $p_T$ in the range 
of 1-4 GeV, the calculations for each individual $d \sigma /dp_T $ 
were done as for $J/\psi$-production 
by incorporating the experimental $x_F$-cut.  The results for differential
cross sections were then fitted as a power law in $A$ to obtain $\alpha(p_T)$.
One sees a similar general agreement for the gas model and the six-quark
cluster model as for $J/\psi$-production at moderate values of $p_T$.  
At the largest $p_T$, however, the E772-data rise too sharply compared to any 
model and could possibly
indicate that these models tuned to earlier large $x$-data have to be
better tuned to perform well in the small $x$-region. 
Fig. \ref{upsalra} shows that this disagreement at large $p_T$, as
well as the agreement at lower $p_T$, are once again features which
do not depend on the hadronisation model.  The figure shows the ratio
of the results obtained for the SLD-model and the CS-model for the   
experimental cuts of E772 as a function of $p_T$ and they are within 
$\sim 5 \%$ of each other.  Thus within the framework of pQCD these data
too are explicable in terms of changes of nuclear structure functions.
As one can expect, the contribution of the $qg$ subprocess in the case
of $\Upsilon$--production is even larger, being typically $\sim$40\% in
comparison to that of the $gg$ subprocess which almost accounted for
the rest.

\section {Dimuon Production}

The $J/\psi (\Upsilon)$ is detected most efficiently via its decay into a
$\mu^+\mu^-$ pair.  Hence any critical evaluation of $J/\psi$
suppression as a signal of QGP formation also needs a good understanding of
the $p_T$- and $A$-dependence of this continuum
$\mu^+\mu^-$ background.  An experimental measurement of the
$A$-dependence of the dimuon pair production can also provide an
independent probe of the nuclear parton densities and an evidence in
favour of the universality of the EMC-effect, {\em i.e.}, its process
independence.  In fact, one of the earliest theoretical attempts to
understand the dimuon data had postulated \cite{rokvl} A-dependent
sea density even before the EMC effect was experimentally discovered. 
The production of massive $\mu^+\mu^-$ pairs (DY) in hadronic collisions
is now well understood in the framework of pQCD\cite{dimuth}.  
The dimuon production at small $p_T$ and large $x_F$ 
is essentially well described in terms of $q\bar q$
annihilation process in spite of the large higher order corrections.
For large $p_T$ of the $\mu^+\mu^-$ pair the production cross-section
is given by the $O(\alpha_s)$ subprocesses involving gluons viz.,
\be \label{hiptmu}
q + g \rightarrow \gamma^\ast + q \rightarrow \mu^+\mu^- + q; ~~ q + \bar q
\rightarrow \gamma^\ast + g \rightarrow \mu^+\mu^- + g
\ee
The high statistics E772-experiment\cite{E7723} has provided data
for nuclear-dependence
of proton-induced pair production over a wide range of $x_F$ and
$p_T$ values.  The data on the ratio of the {\em integrated} dimuon
yield for different nuclei were compared with theoretical predictions,
obtained by using the $q\bar q$ annihilation process, for various models of
the EMC effect.  It seemed\cite{E7723} to rule out the
6-quark cluster model\cite{bag}. 
However, a later comparison\cite{sulassila} with an
improved version of the model, described in sect. II.C,
showed that this model too can be
consistent with the information on the ratio of the integrated dimuon
yields.  The E772 experiment \cite{E7723} has also presented $p_T$
distributions (integrated over $x_F$ and $M^2_{\mu^+\mu^-}$) and $x_F$
distributions (integrated over $p_T$ and $M^2_{\mu^+\mu^-}$) for the
dimuon pairs.  A comparison of ratios of these differential
distributions for different targets with the predictions of various
models of EMC effect can discriminate between them more effectively.
Since the $O(\alpha_s)$ pQCD calculation is valid only at large-$p_T$,
we restrict ourselves to the $p_T$ distributions.  The $x_F$
distributions are integrated over the complete range of $p_T$ and
hence dominated by $p_T \sim 0$ data, which once again forced
us to ignore them in this leading order pQCD analysis.

The kinematics of the DY $\mu^+\mu^-$ pair production at large $p_T$ in
the $O(\alpha_s)$ subprocesses of eq.\,(\ref{hiptmu}) is very similar to the
kinematics of the $J/\psi (\Upsilon)$ production in the colour-singlet model,
discussed in sect. III.B.  Its differential cross-section is given by 
\bea \label{hiptdif}
{d\sigma \over dM^2_{\mu^+\mu^-} dp^2_T} & = & \int^{y^{\rm
max}_1}_{y^{\rm min}_1} dy_1 \int^{y^{\rm max}_2}_{y^{\rm min}_2} dy_2
 ~x_1x_2 \Big\{ P_{qg}~ {d\hat \sigma (qg \rightarrow \mu^+\mu^- q)
\over dM^2_{\mu^+\mu^-} d\hat t} \nonumber\\[2mm]
& + & P_{q\bar q}~{d\hat \sigma \over dM^2_{\mu^+\mu^-} d\hat t}~ (q\bar
q \rightarrow \mu^+\mu^- g) \Big\} ~~,~~
\eea
where
\bea \label{hiptden}
P_{qg}&=&\sum_{q, \bar q} \left[ f_{q/p} (x_1) f_{g/A} (x_2) + f_{g/p}
(x_1) f_{q/A} (x_2) \right] \nonumber\\[2mm]
P_{q\bar q}&=&\sum_q \left[ f_{q/p} (x_1) f_{\bar q /A} (y_2) + f_{\bar
q /p} (x_1) f_{q/A} (x_2) \right]~~,~~
\eea
with
\bea \label{hiptals}
{d\hat\sigma \over d\hat t}~ (qg \rightarrow \mu^+\mu^- q)&=&{\alpha_s
\alpha^2_{em} e^2_q \over 9 M^2_{\mu^+\mu^-}} \cdot {\left\{
\left(\hat s - M^2_{\mu^+\mu^-}\right)^2 + \left(u -
M^2_{\mu^+\mu^-}\right)^2 \right\} \over -\hat s^3 \hat u} \nonumber\\[2mm]
{d\hat \sigma \over d\hat t} ~ (q\bar q \rightarrow \mu^+\mu^- g)&=&
{8~e^2_q \alpha_s \alpha^2_{em} \over 27~M^2_{\mu^+\mu^-}}~ {\left\{
\left(\hat t - M^2_{\mu^+\mu^-}\right)^2 + \left(\hat u -
M^2_{\mu^+\mu^-}\right)^2 \right\} \over \hat s^2 \hat t \hat u}~~.~~
\eea
Here $y_1, y_2$ are the rapidities at the $\mu^+\mu^-$ pair
and the associated jet respectively.  The Mandestam variables $\hat s$,
$\hat t$ and $\hat u$, the relation of the momentum fractions $x_1$ and
$x_2$ in terms of $y_1, y_2$, the integration limits and their relation
with the experimental $x_F$-cut ($x_F > 0$) are precisely the same
as those given in sect. III.B for the colour singlet model with 
$M^2$ replaced in all the formulae by $M^2_{\mu^+\mu^-}$.  
We will therefore not repeat them here.

Experimental information \cite{E7723} is available for the ratio
\be \label{rdyexpt}
R^{DY} = {{d\sigma^{DY} \over dp_T} ~ (p A \rightarrow \mu^+\mu^- 
X) \biggm/ {d\sigma^{DY} \over dp_T} ~ (p p \rightarrow \mu^+\mu^- X)} 
\ee
where ${d\sigma^{DY}/dp_T}$ is the differential $DY$
cross section integrated over the continuum region (avoiding the
resonances) $4 < M_{\mu^+ \mu^-} < 9~{\rm GeV}$ and $M_{\mu^+\mu^-}
\geq 11~ {\rm GeV}$, with $x_F > 0$.  

Integrating eq.\,(\ref{hiptdif}) over the above experimental cut of
$x_F > 0$ and $4 < M_{\mu^+\mu^-} < 9$ as well as $M_{\mu^+\mu^-} >
11~{\rm GeV}$, we compute the ratio $R^{DY}$ of eq.\,(\ref {rdyexpt}) 
for all the different nuclear targets used
in the experiment for each of the three sets of nuclear and nucleon
parton densities described before.  $M^2_{\mu^+\mu^-} $ was used as
the scale for $\alpha_s$ in eqs.\,(\ref{hiptals}).

Fig. \ref{dimu} exhibits the results of our computation for the four different
nuclei with the corresponding data.  Again we see similar to the case
of resonance production that the general trends of the data are well
described by the model predicltions for the gas model and the 6-quark
cluster model.

\section {Conclusion}
 
In conclusion, we have shown in this paper that the high statistics data 
E772 on the nuclear dependence of the production of quarkonia ($J/\Psi$ 
and $\Upsilon$) {\it and} dimuon pairs at large $p_T$ 
can be entirely explained in terms of  the  same nuclear structure
functions, in the frameowrk of pQCD.   All our theoretical calculations
contained no arbitrary free parameters; only existing models of the
EMC-effect with their already fixed values of parameters were used.
We employed two popular models of hadronization of the $ Q \bar Q$-pair
into the quarkonia. In spite of their big kinematical differences, we
found both to yield predictions which were within $\sim 5$\% of each
other.  This shows the robustness of the pQCD approach and 
underlines the importance of the nuclear structure function effects
in understanding the behaviour of these data.  Similar conclusions\cite
{souraj,sriro} about the independence of hadronization mechanism and 
the universality of the nuclear structure functions in
various hard scattering processes\cite{roreview} have been 
obtained before but the accuracy of
the present data makes them now much stronger. Recently it was
argued \cite{helnew} that quantum mechanical coherence and interference 
effects destroy factorisation in quarkonia production and hence
prevent the possibility of using the same nuclear structure functions
for different final states. However, the consistency of both the
quarkonia and the dimuon data with our calculations, points towards
the correctness of ideas of universality of the structure functions,
at least in the kinematic region probed in our analysis. 

Our analysis also indicates that
the accuracy of the data has now reached a stage so as to distinguish
between various models of the EMC effect and the nuclear structure
function parametrisations therein.  Indeed, the inability of
{\em all} models to make a better prediction for the tungsten nucleus
than shown in Fig. 4, and the disagreement at the largest $p_T$ value
in Fig. 6, are hints for inventing better parametrizations of the
nuclear dependence of the quark (and gluon) distributions.  We also
argued that the twin assumptions of gluon dominance and adequacy
of the lowest order partonic cross section are unreliable due to
the large $x_F$ and $p_T$ ranges of the E772-data.  Extraction
of the nuclear gluon density using them is likely to be dominated
by uncertainties as large as the gluonic EMC-effect itself.

We have examined here data at a fixed value of $p_T$ but which have been
integrated over the entire $x_F$ region corresponding to the
acceptance of the experiment. The integrated data are dominated by
the data at small $x_F$ values ( or not-so-small $x_2$ values). Hence our
non-inclusion of any shadowing effects for the nuclear parton densities
can be justified. If one wants to critically use these data to study the
shadowing effects in the nuclear parton densities, then it would be
necessary to look at the nuclear effects in the $p_T$ integrated data
at large values of $x_F$ (which will probe small values of $x_2$).
The data on $x_F$ distributions available currently is integrated
over the entire range of $p_T$ whereas our  pQCD analysis is valid
only for $p_T \ge 1 $ GeV. If information about the $x_F$ distributions
integrated over only the large $p_T$ region becomes available, it will help
unravel the issue of nulear dependence of the $J/\Psi$, $\Upsilon$
and the dimuon production further.

\begin{table}
\caption{ 
Model parameters for nuclei used in E772 experiment for gas model(T,$\omega$) 
and six quark cluster model($\epsilon$).}
\medskip
\begin{tabular}{cccc}
A & T (MeV) & $\omega$ & $\epsilon$ \\
\tableline
12 & 54 & 0.069 & 0.112 \\
40 & 47 & 0.057 & 0.170 \\
56 & 45 & 0.117 & 0.186 \\
184 & 42 & 0.132 & 0.230 \\
\end{tabular}
\label{tabgas}
\end{table}
\medskip
\begin{table}
\caption{ 
Resonance parameters used for $\psi,\psi^{'}$ and $\Upsilon, \Upsilon^{'}$.}
\medskip
\begin{tabular}{cccc}
Resonance R&M(MeV)&$R(0)^2 (GeV)^3$ &BF$(R \rightarrow ~^3S_1 + $ neutrals) \\
\tableline
$\psi $ & 3096 & 0.542 & 1.00\\
$\psi^{'}$ & 3686 & 0.307 & 0.55 \\
$\Upsilon$ & 9460 & 4.54  & 1.00 \\
$\Upsilon^{'}$ & 10020 & 2.54 & 0.19 \\
\end{tabular}
\label{tabres1}
\end{table}
\medskip
\begin{table}
\caption{ 
Resonance parameters used for $\chi^{c}$ and $\chi^{b}$ states.}
\medskip
\begin{tabular}{cccc}
Resonance R&M(MeV)&$R_1^\prime(0)^2/M^2 (GeV)^3$ &BF$(R \rightarrow
^3S_1 + $ neutrals ) \\
\tableline
$\chi_0^c $ & 3415 &$9.1 \times 10^{-3}$  & $6.6 \times 10^{-3}$ \\
$\chi_1^c $ & 3510 &$9.1 \times 10^{-3}$  & 0.273 \\
$\chi_2^c $ & 3555 &$9.1 \times 10^{-3}$  & 0.135 \\
\hline
$\chi_0^b $ & 9860 &$1.5 \times 10^{-2}$  & 0.040 \\
$\chi_1^b $ & 9890 &$1.5 \times 10^{-2}$  & 0.290 \\
$\chi_2^b $ & 9915 &$1.5 \times 10^{-2}$  & 0.220 \\
\end{tabular}
\label{tabres2}
\end{table}
\clearpage
\begin{figure}
%\narrowtext
\caption{Predictions for the ratio of the nuclear and nucleonic gluon
density, $\rho_g$, of the three models of the EMC
effect described in the text.\label{fig1}}
\end{figure}

\begin{figure}
\noindent
\caption{The ratio $R_{qg} = {{F_{qq}} \over {F_{gg}}}$ of Eq. (28)
corresponding to $J/\psi$  production at the FNAL energies for
MT1(solid), MT2(long--dashed), MT3(medium--dashed), ON(short--dashed),
GGR(dotted), GHR(dot--dashed), DO1(dot--dot--dashed) and DO2(inverted
triangle) parametrisation of the parton densities.\label{fig2}}
\end{figure}

\begin{figure}
\caption{Same as in Fig. \protect\ref{fig2} but for Upsilon
production at the FNAL energies.~~~~~~~~~~~~~~~~~~\label{fig3}}
\end{figure}

\begin{figure}
\caption{E772 data on the ratio $R^{J/\psi}$ of Eq. 29 compared with the
predictions  for the gas model(squares), six-quark cluster
model(circles) and the rescaling model(open circles) of the EMC
effect, obtained using the SLD for hadronisation.}\label{jpsisld}
\end{figure}

\begin{figure}
\caption{Ratio of the predictions for $R^{J/\psi}$ for the
SLD and CS model of hadronisation, for the
different nuclear parton densities. Notation is same as in Fig. 
\protect\ref{jpsisld}.} \label{jpsirat}
\end{figure}

\begin{figure}
\caption{E772 data on $\alpha (p_T)$ of Eq. 30 compared with
predictions of the three different models of the EMC effect 
mentioned in the text, using the SLD model of hadronisation. 
Notation is same as in Fig. \protect\ref{jpsisld}.} \label{upsalph}
\end{figure}

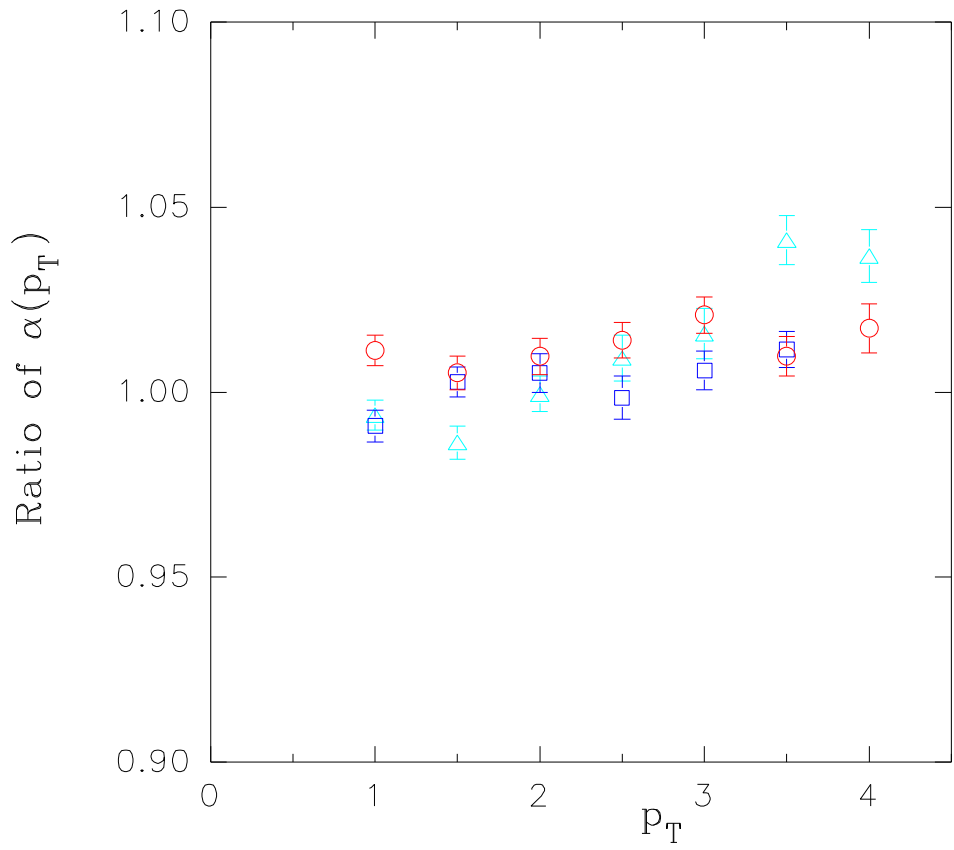
\begin{figure}
\caption{Ratio of $\alpha (p_T)$ predicted in the SLD and CS models of
hadronisation for the three models of the EMC effect. Notation is the
same as in Fig. \protect\ref{jpsisld}.} \label{upsalra}
\end{figure}

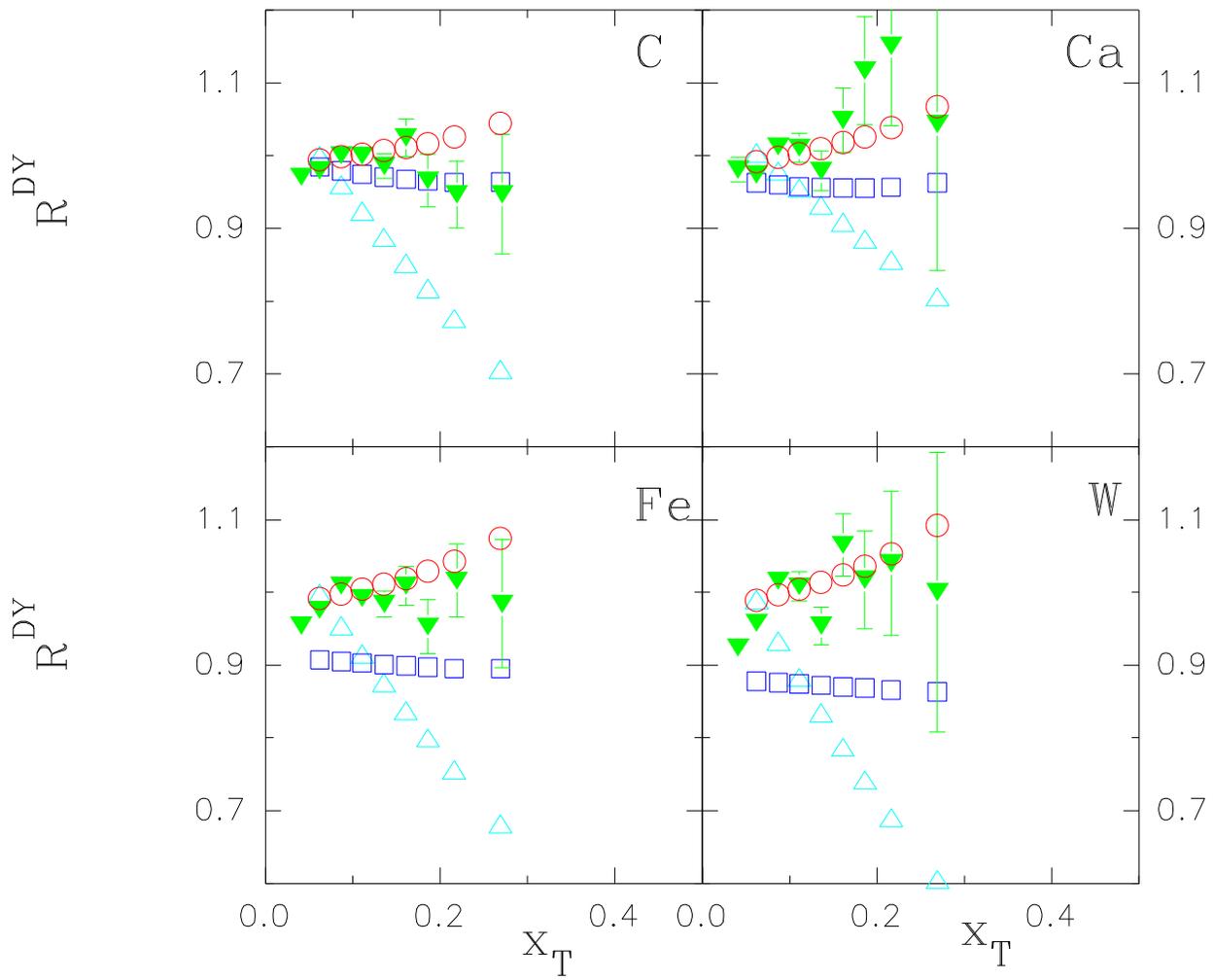
\begin{figure}
\caption{E772 data on the ratio $R^{DY}$ of eq.\protect\ref{rdyexpt}
compared with predictions of the three models of the EMC effect. 
Notation is same as in Fig. \protect\ref{jpsisld}.} \label{dimu}
\end{figure}
\end{document}